\documentclass[aps,prl,twocolumn,groupedaddress,showpacs]{revtex4}
\usepackage{graphicx}
\begin{document}

\title{Spin quantum tunneling via entangled states in a dimer of exchange
coupled single-molecule magnets}

\author{R. Tiron$^1$, W. Wernsdorfer$^1$, D. Foguet-Albiol$^2$, 
N. Aliaga-Alcalde$^2$, G. Christou$^2$}

\affiliation{
$^1$Lab. L. N\'eel, associ\'e \`a l'UJF, CNRS, BP 166,
38042 Grenoble Cedex 9, France\\
$^2$Dept. of Chemistry, Univ. of Florida,
Gainesville, Florida 32611-7200, USA
}

\date{18th July 2003, submitted to PRL}

\begin{abstract}
A new family of supramolecular, antiferromagnetically 
exchange-coupled dimers of single-molecule magnets 
(SMMs) has recently been reported 
[W. Wernsdorfer, N. Aliaga-Alcalde, 
D.N. Hendrickson, and G. Christou, Nature 416,  406  (2002)]. 
Each SMM acts as a bias on its neighbor, 
shifting the quantum tunneling resonances 
of the individual SMMs. Hysteresis loop 
measurements on a single crystal of SMM-dimers 
have now established quantum tunneling of 
the magnetization via entangled states of the dimer. 
This shows that the dimer really does behave 
as a quantum-mechanically coupled dimer. 
The transitions are well separated, suggesting 
long coherence times compared to the time scale 
of the energy splitting. This result is of great 
importance if such systems are to be used for 
quantum computing. It also allows the measurement 
of the longitudinal and transverse superexchange coupling constants.
\end{abstract}

\maketitle

Single-molecule magnets (SMM) are among the 
smallest nanomagnets that exhibit 
magnetization hysteresis, a classical property of macroscopic 
magnets~\cite{Christou00,Sessoli93b,Sessoli93,Aubin96,Boskovic02}. 
They straddle the interface 
between classical and quantum mechanical behavior because they also 
display quantum tunneling of magnetization
~\cite{Novak95,Friedman96,Thomas96,Sangregorio97,Hill98,Aubin98,Kent00b,WW_Mn30} 
and quantum phase interference~\cite{WW_Science99,Garg93}. 
These molecules comprise several
magnetic ions, whose spins are coupled by strong exchange interactions
to give a large effective spin. The molecules are regularly
assembled within large crystals, with all the molecules often having the same
orientation. Hence, macroscopic measurements can give direct access
to single molecule properties. Many non-magnetic atoms surround the
magnetic core of each molecule. Exchange interactions between 
molecules are therefore relatively
weak and have been neglected in most studies.
Recently, the study of a dimerized SMM, [Mn$_4$]$_2$, showed
that intermolecular exchange interactions are not always
negligible and can instead be used to couple SMMs. This 
system~\cite{WW_Nature02} was 
[Mn$_4$O$_3$Cl$_4$ (O$_2$CEt)$_3$(py)$_3$]$_2$ 
(hereafter called [Mn$_4$]$_2$), a member of the  
[Mn$_4$O$_3$Cl$_4$(O$_2$C{\bf R})$_3$(py)$_3$]$_2$
family, with {\bf R} =Et . The crystal form studied, 
[Mn$_4$]$_2\cdot$8MeCN, was that obtained from 
MeCN solution, containing 4 MeCN solvent molecules 
of crystallization per Mn$_4$. 
The spins of the two Mn$_4$ molecules are
coupled antiferromagnetically. Each molecule acts as a bias on its neighbor,
the quantum tunneling resonances thus being shifted with 
respect to the isolated Mn$_4$
SMM. The first three-dimensional networks of
exchange coupled SMMs have also been studied recently~\cite{Yang03,Tiron03}.

In this letter, we present new results discovered on a different 
crystal form of the same [Mn$_4$]$_2$ compound,  
obtained from CH$_2$Cl$_2$/Et$_2$O/C$_6$H$_{14}$ 
solution and containing one hexane (C$_6$H$_{14}$) 
molecule of crystallization per Mn$_4$ i.e. [Mn$_4$]$_2\cdot$2C$_6$H$_{14}$. 
Both [Mn$_4$]$_2\cdot$8MeCN and [Mn$_4$]$_2\cdot$2C$_6$H$_{14}$ 
crystallize isomorphously, but the latter has a 
stronger intradimer and negligible interdimer
exchange interactions compared with the former, 
and was thus better suited for the studies presented here.
We have identified for the first time quantum tunneling transitions 
via entangled states of the [Mn$_4$]$_2$ dimer. 
The corresponding energy levels are well separated, 
showing that the decoherence in this system 
is small.
In our previous report, we did not have, and thus 
could not provide, evidence for quantum mechanical entanglement 
within this dimer, but the present results establish
that the dimer really does behave as a quantum-mechanically coupled system.

The compound [Mn$_4$]$_2\cdot$2C$_6$H$_{14}$ crystallizes in the
hexagonal space group $R3$(bar) with two Mn$_4$
molecules per unit cell lying head-to-head on a crystallographic S$_6$
symmetry axis~\cite{Nuria03}, as does previously 
reported [Mn$_4$]$_2\cdot$8MeCN~\cite{WW_Nature02}. 
Each Mn$_4$ monomer has a ground state spin of
$S$ = 9/2, well separated from the first excited state $S$ = 7/2
by a gap of about 300K~\cite{Hendrickson92}. 
The Mn-Mn distances and the Mn-O-Mn
angles are similar and the uniaxial anisotropy constant is expected to be
the same for the two dimer systems. These dimers are held together via six
C$-$H$\cdot\cdot\cdot$Cl hydrogen bonds between the pyridine (py) rings
on one molecule
and the Cl ions on the other, and one Cl$\cdot\cdot\cdot$Cl Van der 
Waals interaction. These interactions lead to an antiferromagnetic
superexchange interaction between the two Mn$_4$ units of the 
[Mn$_4$]$_2$ dimer~\cite{WW_Nature02}. 
Dipolar couplings between Mn$_4$ molecules can be 
easily calculated and are more than one order 
of magnitude smaller than the exchange interaction.

\begin{figure}
\begin{center}
\includegraphics[width=.42\textwidth]{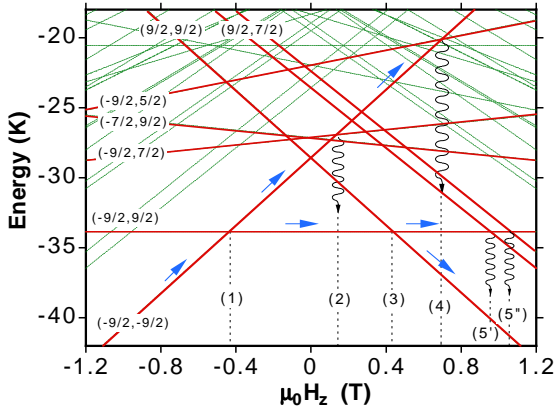}
\caption{(Online color) Low lying spin state energies of
the [Mn$_4$]$_2$ dimer, calculated 
by exact numerical diagonalization using Eq. 2
with $D$ = 0.77 K and $J$ = 0.13 K,
as a function of applied magnetic field $H_z$ (Zeeman diagram).
The bold energy levels are labelled with 
two quantum numbers $(M_1,M_2)$.
Dotted lines, labelled {\bf 1} to {\bf 5}, indicate 
the strongest tunnel resonances: 
{\bf 1}: (-9/2,-9/2) to (-9/2,9/2); 
{\bf 2}: (-9/2,-9/2) to (-9/2,7/2), 
followed by relaxation to (-9/2,9/2); 
{\bf 3}: (-9/2,9/2) to (9/2,9/2); 
{\bf 4}: (-9/2,-9/2) to (-9/2,5/2), followed by relaxation to (-9/2,9/2); 
{\bf 5}: (-9/2,9/2) to (7/2,9/2), followed by relaxation to (9/2,9/2). 
For clarity, degenerate states such as (M,M') and (M',M) 
and lifted degenerate states such as 
$(M,M \pm 1)$, $(M,M \pm 2)$ \ldots are not both listed.
For example,
the $(9/2,7/2)$ and $(7/2,9/2)$ states are strongly
split into a symmetric (labelled {\bf $5''$}) and antisymmetric 
(labelled {\bf $5'$})
combination of $(9/2,7/2)$ and $(7/2,9/2)$ states. 
This splitting is used to measure the
transverse superexchange interaction constant $J_{xy}$.
Co-tunneling and other two-body tunnel transitions 
have a lower probability of 
occurrence and are neglected~\cite{WW_PRL02}.}
\label{fig1}
\end{center}
\end{figure}

Before presenting the measurements, 
we summarize a simplified spin Hamiltonian 
describing the [Mn$_4$]$_2$ dimer~\cite{WW_Nature02}. 
Each Mn$_4$ SMM can be modeled as a {\it giant spin} of $S = 9/2$ 
with Ising-like anisotropy. The corresponding Hamiltonian is given by
\begin{equation}
	\mathcal{H}_i = -D S_{z,i}^2 + \mathcal{H}_{{\rm trans}, i} 
	+ g \mu_{\rm B} \mu_0 \vec{S_i}\cdot\vec{H} 
\label{eq_Hi}
\end{equation}
where $i = 1$ or 2 (referring to the two Mn$_4$ SMMs of the dimer), 
$D$ is the uniaxial anisotropy constant, 
and the other symbols have their usual meaning. 
Tunneling is allowed in these half-integer ($S = 9/2$) 
spin systems because of a small transverse anisotropy $\mathcal{H}_{{\rm trans}, i}$ 
containing $S_{x,i}$ and $S_{y,i}$  spin operators and transverse 
fields ($H_{x}$ and $H_{y}$). 
The exact form of $\mathcal{H}_{{\rm trans}, i}$  is not important 
in this discussion. The last term in Eq. 1 is the 
Zeeman energy associated with an applied field. 
The Mn$_4$ units within the [Mn$_4$]$_2$ dimer 
are coupled by a weak superexchange interaction via 
both the six C-H$\cdot\cdot\cdot$Cl pathways 
and the Cl$\cdot\cdot\cdot$Cl approach. 
Thus, the Hamiltonian ($\mathcal{H}$) for [Mn$_4$]$_2$ is
\begin{equation}
	\mathcal{H} = \mathcal{H}_1 + \mathcal{H}_2 
	+ J_z S_{z,1}S_{z,2} 
	+ J_{xy} (S_{x,1}S_{x,2} + S_{y,1}S_{y,2})
\label{eq_H}
\end{equation}
where $J_z$ and $J_{xy}$ are respectively the longitudinal 
and transverse superexchange interactions.  $J_z = J_{xy}$
is the case of isotropic superexchange. 
The $(2S+1)^2 = 100$ energy states of the dimer 
can be calculated by exact numerical diagonalization and are plotted 
in Fig. 1 as a function of applied field along the easy axis. 
Each state of [Mn$_4$]$_2$ can be labelled by two quantum 
numbers $(M_1,M_2)$ for the two Mn$_4$ SMMs, 
with $M_1 = -9/2, -7/2, ..., 9/2$ and $M_2 = -9/2, -7/2, ..., 9/2$. 
The degeneracy of some of the $(M_1,M_2)$ states
is lifted by transverse anisotropy terms. For the sake of simplicity,
we will discuss mainly the effect of 
the transverse superexchange interaction
$\mathcal{J}_{trans} = J_{xy} (S_{x,1}S_{x,2} + S_{y,1}S_{y,2}) = 
J_{xy} (S_{+,1}S_{-,2} + S_{-,1}S_{+,2})/2$,
where $S_{+,i}$ and $S_{-,i}$ are the usual 
spin raising and lowering operators.
Because $\mathcal{J}_{trans}$ acts on $(M,M \pm 1)$ states
to first order of perturbation theory, the degeneracy
of those states is strongly lifted. For example,
the $(9/2,7/2)$ and $(7/2, 9/2)$ states are strongly
split into a symmetric (labelled {\bf $5''$}) and antisymmetric 
(labelled {\bf $5'$})
combination of $(9/2,7/2)$ and $(7/2,9/2)$ states. 
Similarly for the (-9/2, -7/2) and (-7/2, -9/2) states.
Measuring this energy splitting allows us to determine the
transverse superexchange interaction constant $J_{xy}$
because the latter is proportional to the former.

\begin{figure}
\begin{center}
\includegraphics[width=.42\textwidth]{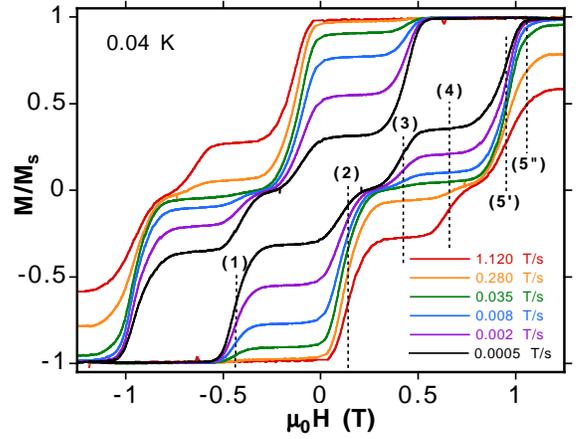}
\caption{(Online color) Hysteresis loops for the [Mn$_4$]$_2$ dimer
at several field sweep rates and 40 mK.
The tunnel transitions (manifested by steps) 
are labelled from {\bf 1} to {\bf 5},
see Fig. 1.}
\label{fig2}
\end{center}
\end{figure}

Tunneling studies on [Mn$_4$]$_2$ were performed 
by magnetization measurements on single crystals 
using an array of micro-SQUIDs~\cite{WW_ACP_01}. 
Fig. 2 shows typical hysteresis loops (magnetization versus
magnetic field scans) with the field applied along 
the easy axis of magnetization of [Mn$_4$]$_2$, that is, parallel to the 
S$_6$ axis.
These loops display step-like features separated by plateaus. 
The step heights are temperature-independent below
$\sim$0.35 K (not shown). The steps are due to resonant 
quantum tunneling of the magnetization (QTM) between 
the energy states of the [Mn$_4$]$_2$ dimer (see figure 
caption 1 and 2 for a discussion of 5 tunnel transitions). 
QTM has been previously observed for most SMMs,
but the novelty for [Mn$_4$]$_2$ dimers is that the QTM is now 
the collective behavior of the complete $S = 0$ dimer 
of exchange-coupled $S = 9/2$ Mn$_4$ quantum systems. 
This coupling is manifested as an exchange bias 
of all tunneling transitions, and the resulting 
hysteresis loop consequently displays unique features, 
such as the absence for the first time in a SMM of a QTM 
step at zero field~\cite{WW_Nature02}.

Even though the five strongest tunneling transitions are
observed in Fig. 2, fine structure was not
observed. For example, the hysteresis loops
do not show the
splitting of the $(9/2,7/2)$ states (labelled {\bf $5'$}
and {\bf $5''$}), which we suspected might be due to 
line broadening. Usually, line broadening in SMMs 
is caused by dipolar and hyperfine interactions
~\cite{Prokofev98}, and distributions of
anisotropy and exchange parameters.
In most SMMs, the zero-field resonance is
mainly broadened by dipolar and hyperfine interactions
because distributions of anisotropy parameters
do not affect the zero-field resonance.
For an antiferromagnetically coupled dimer, however, 
this resonance is shifted to negative fields.
Therefore, a distribution of the exchange coupling
parameter $J_z$ can further broaden this resonance.
In fact, we show in the following that the latter is the 
dominant source of broadening.
We then use the `quantum hole-digging' method 
~\cite{Prokofev98,WW_PRL99,WW_PRL00,Fernandez01,Tupitsyn03}
to provide direct experimental evidence for the transitions {\bf $5'$} and {\bf $5''$},
which establishes tunneling involving entangled dimer states
and allows us to determine $J_{xy}$.

\begin{figure}
\begin{center}
\includegraphics[width=.42\textwidth]{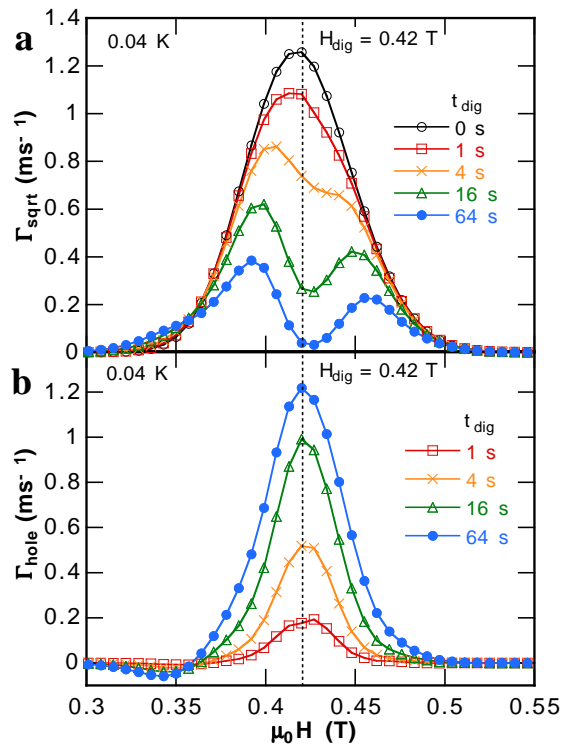}
\caption{(Online color) (a) Field dependence of the
short-time square-root relaxation rates $\Gamma_{\rm sqrt}$
are presented on a logarithmic scale showing 
the depletion of the molecular
spin states by quantum tunneling at $H_{\rm dig} = 0.42$ T
for various waiting times $t_{\rm dig}$. 
(b) Difference between the relaxation rate in the
absence and in the presence of digging,
$\Gamma_{\rm hole} =
\Gamma_{\rm sqrt}(t_{\rm dig}=0) -
\Gamma_{\rm sqrt}(t_{\rm dig})$.}
\label{fig3}
\end{center}
\end{figure}

The `quantum hole-digging' method
is a relatively new method that can, 
among other things~\cite{WW_Mn30}, 
study line broadening and its evolution
during relaxation~\cite{Prokofev98,WW_PRL99,WW_PRL00,Fernandez01,Tupitsyn03}.
The method is based on the simple idea that 
after a rapid field change, the resulting 
magnetization relaxation at short time periods 
is directly related to the number of molecules 
in resonance at the applied field; 
Prokof'ev and Stamp proposed~\cite{Prokofev98} that this 
short time relaxation should follow a $\sqrt{t}$ 
($t$ = time) relaxation law. Thus, the magnetization 
of the [Mn$_4$]$_2$ dimers in the crystal was first 
saturated with a large positive field, and then a 
`digging field' $H_{\rm dig}$ was 
applied at 0.04 K for a chosen `digging time' $t_{\rm dig}$. 
Then, the fraction (and only that fraction) 
of the molecules that 
is in resonance at $H_{\rm dig}$ can undergo 
magnetization tunneling. After $t_{\rm dig}$, a field 
$H_{\rm probe}$ is applied and the magnetization 
relaxation rate is measured for short time periods; 
from this is calculated the short-time relaxation 
rate $\Gamma_{\rm sqrt}$, which is related to the number 
of [Mn$_4$]$_2$ dimers still available for QTM~\cite{WW_ACP_01}. 
The entire procedure is then repeated at 
other $H_{\rm probe}$ fields. 
The resulting plot (Fig. 3a) of $\Gamma_{\rm sqrt}$ versus $H_{\rm probe}$ 
reflects the distribution of spins still available 
for tunneling after $t_{\rm dig}$. 

\begin{figure}
\begin{center}
\includegraphics[width=.42\textwidth]{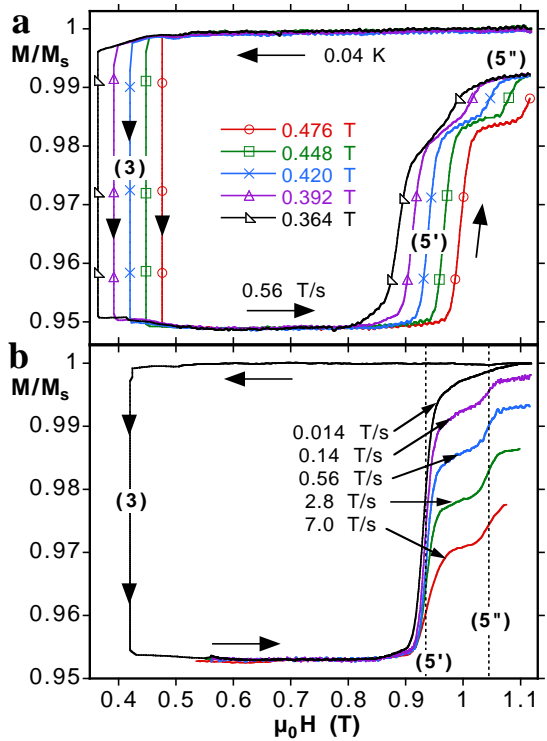}
\caption{(Online color) Minor hysteresis loops for 
several (a) digging fields and (b) field sweep rates. 
After positive saturation, a 
digging field $H_{\rm dig}$ was 
applied to reverse $\approx$2.5\% of the molecules that 
are in resonance at $H_{\rm dig}$ (transition {\bf 3} in Fig. 1). 
Then, the applied field is swept back to a large positive field.
{\bf $5'$} and {\bf $5''$} are the first tunnel transitions 
allowing the reversed molecules to tunnel back to positive saturation.}
\label{fig4}
\end{center}
\end{figure}

In the limit of very short digging times, the difference between the
relaxation rate in the
absence and in the presence of digging,
$\Gamma_{\rm hole} =
\Gamma_{\rm sqrt}(t_{\rm dig}=0) -
\Gamma_{\rm sqrt}(t_{\rm dig})$, is
approximately  proportional to the number of molecules which reversed their
magnetization during
the time $t_{\rm dig}$ (Fig. 3b). 
$\Gamma_{\rm hole}$ is characterized
by a width that can be called the 
`tunnel window'.

The width of the distribution in the absence of digging
($\sim$80 mT, Fig. 3a) is too large to be due to only
dipolar ($\sim$20 mT) and hyperfine coupling ($\sim$10 mT).
The following result suggests that it is due to a 
distribution of the exchange coupling parameter $J_z$.

\begin{figure}
\begin{center}
\includegraphics[width=.42\textwidth]{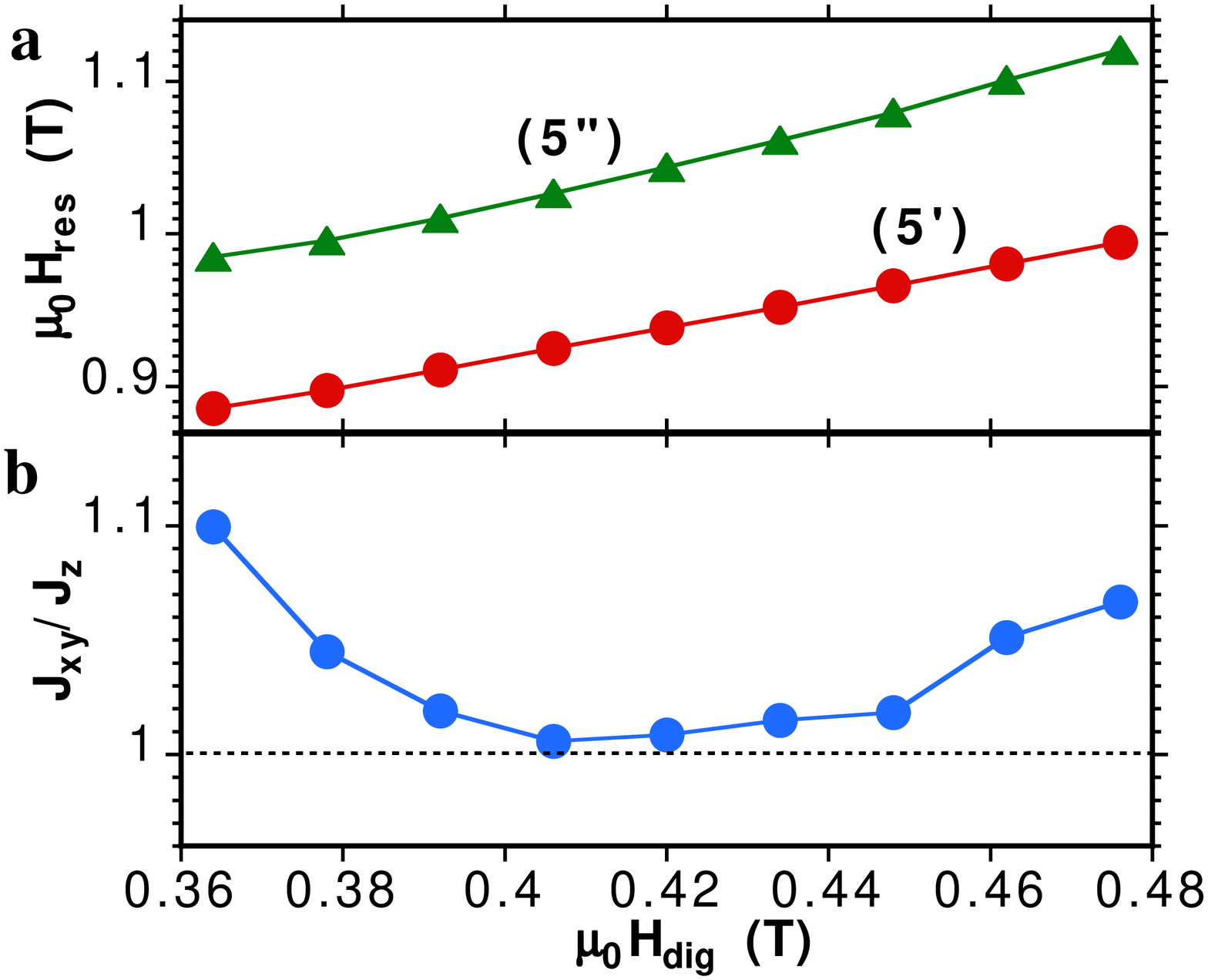}
\caption{(Online color) (a) Resonance field positions $H_{\rm res}$ 
of {\bf $5'$} and {\bf $5''$} and (b) normalized 
transverse superexchange interaction $J_{xy}/J_z$ as a
function of digging field $H_{\rm dig}$.}
\label{fig5}
\end{center}
\end{figure}

First, the magnetization of the [Mn$_4$]$_2$ 
dimers was saturated with a large positive field, and then a 
`digging field' $H_{\rm dig}$ was 
applied to reverse a fraction of the molecules that 
are in resonance at $H_{\rm dig}$ (transition {\bf 3} in Fig. 1). 
After the reversal of 2.5\% of the molecules,
the applied field is swept back to a large positive field.
{\bf $5'$} and {\bf $5''$} are the first tunnel transitions 
that can allow the reversed molecules
to tunnel back to positive saturation.
Figs. 4a and 4b show the corresponding minor hysteresis
loops for several 'digging fields and
field sweep rates, respectively. Both figures
show clearly the expected tunnel transitions {\bf $5'$} and {\bf $5''$},
that were not resolved in the major hysteresis loops (Fig. 2).
This suggests that the broadening of tunneling transition {\bf 3}
(the distribution in the absence of digging in Fig. 3a) 
is dominated by a 
distribution of the exchange coupling parameter $J_z$.
During the application of the digging field, a subgroup
of molecules is selected with an exchange coupling 
constant $J_z \approx g \mu_{\rm B} \mu_0 H_{\rm dig}/S$,
that can tunnel back at the fields of transitions {\bf $5'$} and {\bf $5''$}.

This interpretation is supported by the study of
the field values of {\bf $5'$} and {\bf $5''$} as a
function of digging field, that is $J_z$,
exhibiting a nearly linear variation (Fig. 5a). 
The field difference between transition {\bf $5'$} and {\bf $5''$}
can be used to find the $J_{xy}$, presented in Fig. 5b.
This shows that the superexchange interaction 
of the dimers is nearly isotropic ($J_{xy} \approx J_{z}$).
It is important to mention that the transitions {\bf $5'$} and {\bf $5''$}
are well separated, suggesting long coherence times
compared to the time scale of the energy splitting.

The above results demonstrate for the first time
tunneling via entangled states of a dimer
of exchange coupled SMMs, showing 
that the dimer really does behave 
as a quantum mechanically coupled system. 
This result is of great 
importance if such systems 
are to be used for quantum computing.

We thank the CNRS, Rhone-Alpe, 
and USA National Science Foundation 
for support.


\end{document}